\documentclass{ieeeaccess}
\usepackage[utf8]{inputenc}
\usepackage{cite}
\usepackage{amsmath,amssymb,amsfonts}
\usepackage{algorithmic}
\usepackage{graphicx}
\usepackage{tikz}
\NewSpotColorSpace{PANTONE}
  \AddSpotColor{PANTONE} {PANTONE3015C} {PANTONE\SpotSpace 3015\SpotSpace C} {1 0.3 0 0.2}
  \SetPageColorSpace{PANTONE}%
\usepackage{textcomp}
\usepackage{soul}
\usepackage{multirow}
\usepackage{algorithm}
\usepackage{algorithmic}
\usepackage{textcomp}
\def\BibTeX{{\rm B\kern-.05em{\sc i\kern-.025em b}\kern-.08em
    T\kern-.1667em\lower.7ex\hbox{E}\kern-.125emX}}
\usepackage{amsthm}
\usepackage{ bbold }

\usepackage{scalerel}
\usepackage{tikz}
\usetikzlibrary{svg.path}

\definecolor{orcidlogocol}{HTML}{A6CE39}
\tikzset{
  orcidlogo/.pic={
    \fill[orcidlogocol] svg{M256,128c0,70.7-57.3,128-128,128C57.3,256,0,198.7,0,128C0,57.3,57.3,0,128,0C198.7,0,256,57.3,256,128z};
    \fill[white] svg{M86.3,186.2H70.9V79.1h15.4v48.4V186.2z}
                 svg{M108.9,79.1h41.6c39.6,0,57,28.3,57,53.6c0,27.5-21.5,53.6-56.8,53.6h-41.8V79.1z M124.3,172.4h24.5c34.9,0,42.9-26.5,42.9-39.7c0-21.5-13.7-39.7-43.7-39.7h-23.7V172.4z}
                 svg{M88.7,56.8c0,5.5-4.5,10.1-10.1,10.1c-5.6,0-10.1-4.6-10.1-10.1c0-5.6,4.5-10.1,10.1-10.1C84.2,46.7,88.7,51.3,88.7,56.8z};
  }
}

\newcommand\orcidicon[1]{\href{https://orcid.org/#1}{\mbox{\scalerel*{
\begin{tikzpicture}[yscale=-1,transform shape]
\pic{orcidlogo};
\end{tikzpicture}
}{|}}}}

\usepackage{hyperref} 

\usepackage{booktabs}
\usepackage{soul}

\usepackage{amsfonts}

\usepackage{multirow}

\def\mbf#1{\mbox{\boldmath$#1$}}

\begin{document}

\title{Self-supervised deep convolutional neural network for chest X-ray classification}

\author{\uppercase{Matej Gazda}\authorrefmark{1,2}\orcidicon{0000-0002-8405-7017},
\uppercase{J\'an Plavka\authorrefmark{2}\orcidicon{0000-0002-2728-108X} , Jakub Gazda\orcidicon{0000-0003-2246-2217}
}\authorrefmark{3}, and \uppercase{Peter Drot\'ar}\authorrefmark{1} \orcidicon{0000-0002-6634-4696} 
\IEEEmembership{Member, IEEE}}
\address[1]{Intelligent Information Systems Lab, Faculty of Electrical Engineering and Informatics, Technical University of Kosice, Letna 9,  04201 Kosice, Slovak Republic (e-mail: matej.gazda@tuke.sk, peter.drotar@tuke.sk)}
\address[2]{Department of Mathematics and Theoretical Informatics, Faculty of Electrical Engineering and Informatics, Technical University of Kosice, Letna 9,  04201 Kosice, Slovak Republic (e-mail: jan.plavka@tuke.sk).}
\address[3]{2nd Department of Internal Medicine, Pavol Jozef Safarik University and Louis Pasteur University Hospital, Trieda SNP 1, 04011, Kosice, Slovak Republic (e-mail: jakub.gazda@upjs.sk).}
\tfootnote{This work was supported by the Slovak Research and Development Agency under contract No. APVV-16-0211 and by the Scientific Grant Agency of the Ministry of Education, Science, Research and Sport of the Slovak Republic and the Slovak Academy of Sciences under contract VEGA 1/0327/20.}

\markboth
{Author \headeretal: Preparation of Papers for IEEE Access Journal}
{Author \headeretal: Preparation of Papers for IEEE Access Journal}

\corresp{Corresponding author: Peter Drotar (e-mail: peter.drotar@tuke.sk).}

\begin{abstract}

Chest radiography is a relatively cheap, widely available medical procedure that conveys key information for making diagnostic decisions. Chest X-rays are frequently used in the diagnosis of respiratory diseases such as pneumonia or COVID-19. In this paper, we propose a self-supervised deep neural network that is pretrained on an unlabeled chest X-ray dataset. Pretraining is achieved through the contrastive learning approach by comparing representations of differently augmented input images. The learned representations are transferred to downstream tasks -- the classification of respiratory diseases. We evaluate the proposed approach on two tasks for pneumonia classification, one for COVID-19 recognition and one for discrimination of different pneumonia types. The results show that our approach yields competitive results without requiring large amounts of labeled training data.

\end{abstract}

\titlepgskip=-15pt
\doi{10.1109/ACCESS.2021.3125324}
\maketitle
\section{Introduction}

Medical imaging utilization has increased rapidly in recent decades, increasing by more than 50\% for some modalities \cite{smith2012}. Although the rate of increase has slowed in recent years \cite{hong2020}, medical imaging is still considered a significant diagnostic method.

Among all medical imaging modalities, radiography is cost effective and is frequently employed by hospitals, emergency services, and other medical facilities. Chest radiography (or chest X-ray (CXR)) is a painless, noninvasive, and powerful investigatory method that conveys crucial respiratory disease information. For respiratory diseases, CXR is a basal diagnostic tool. In CXR, pulmonary opacification (``the white lung field'') is the result of a decrease in the ratio of gas to soft tissue in the lung. Pulmonary opacification has several likely causes, including atelectasis, bronchogenic carcinoma, pleural effusion, tuberculosis, or bacterial or viral pneumonia.

A diagnosis of pneumonia is usually made after considering a combination of clinical symptoms (cough, fever, pathological respiratory sounds), laboratory results (white blood cell count, C-reactive protein and procalcitonin levels, blood gas analysis, and sputum culture), and the presence of pulmonary opacification on CXR. Although the diagnosis and treatment of pneumonia are straightforward in most cases, rapid and accurate diagnosis is specifically required in uncertain cases because complications resulting from an initial misdiagnosis may lead to prolonged hospitalization and resource-draining medical care. Pneumonia is one of the most frequent causes of death in patients of all ages \cite{RUUSKANEN2011} and accounts for a significant number of hospital admissions \cite{Brown2012}.

The recent outbreak of coronavirus disease 2019 (COVID-19) has ushered in unprecedented challenges for most medical facilities. The enormous number of infections calls not only for prevention but also for early diagnosis followed by effective treatment. In this scenario, chest radiography has proven 
to be one of the most time- and cost-effective tools for COVID-19 diagnosis \cite{wongc19}. Upon CXR, patients who suffer from COVID-19 pneumonia present a combination of different multifocal patterns of pulmonary opacification. However, in contrast to community-acquired bacterial pneumonia, these changes are frequently bilateral. Furthermore, while the distribution of these changes is initially peripheral, during the course of the disease, they usually spread to other parts of the lung parenchyma as well. A recent study \cite{Russel2020} showed that the correct diagnosis of mild and moderate COVID-19 from CXR is challenging, even for experienced radiologists. A shortage of medical personnel, the necessity of large numbers of diagnostic decisions even under unfavorable conditions, and the demand for quick and accurate medical decisions all mean that the need for computer-aided diagnostics is greater than ever. %

In this study, we addressed the two most pronounced use cases of the CXR classification. Pneumonia is one of the most frequent and serious inflammatory conditions, and COVID-19 is currently inflicting a devastating impact on healthcare services and even economies in many states worldwide.

This study makes several contributions. First, we trained a deep convolutional neural network (CNN) in a self-supervised fashion using pseudolabels generated through random data augmentation on a large dataset of unlabeled CXR images. This goes beyond current state-of-the art techniques that rely on a large corpus of labeled data. Second, we proposed utilizing this pretrained CNN as a feature extractor for several downstream tasks aimed at CXR classification. Third, even though the proposed CXR classification network does not require large amounts of labeled data, it achieves performance levels competitive to those of supervised counterparts. Extensive experiments on COVID-19 and pneumonia detection tasks validate that the proposed model obtains very reasonable CXR feature representations; thus, it enables accurate CXR classifications on the four evaluated datasets.

The remainder of this paper is organized as follows. In Section II, we provide a brief overview of related works on CNN utilization for CXR classification and contrastive learning. In Section III, we introduce the proposed approach for a self-supervised CNN and its architecture. In Sections IV and V, we describe the datasets used in this study, present the experiments and report their results. Finally, after providing a discussion in Section VI, we draw conclusions in Section VII.

\section{Related works}
In this section, we present related works on CXR classification and the methods involved in our work.

\subsection{Convolutional neural networks for CXR classification}

Two main enablers have emerged in the CXR classification domain in recent years. The most frequent limitation for successful pattern recognition in the biomedical imaging domain has always been (and still is) a scarcity of data. This problem was partially overcome with the introduction of transfer learning for CNNs. Fine-tuned CNNs have shown enormous potential and have even outperformed fully trained CNNs in many applications \cite{tajbakhsh}. The second enabler involves data: several large CXR datasets have been made publicly available, permitting the utilization of many new methodologies for CXR classification \cite{irvin2019chexpert,wang2017chestx}. Most of the forthcoming works have taken advantage of one or both of these enablers.

CXR classification has drawn attention from the research community for several years, but the arrival of the COVID-19 pandemic has boosted interest in this topic. The majority of recent works on CXR classification focus solely or partially on COVID-19 classification.

Given recent findings, the most straightforward approach to diagnosing COVID-19 from CXR images is to use existing CNN architectures pretrained on ImageNet and fine-tune them on a target COVID-19 dataset. This was the approach employed by the authors of \cite{minaee}. They fine-tuned four state-of-the-art convolutional networks (ResNet18, ResNet50, SqueezeNet, and DenseNet-121) to identify COVID-19. Similarly, Apostolopoulos et al. \cite{apostolo} evaluated five other CNN architectures pretrained on ImageNet and found that the most promising results were achieved by the VGG-19 architecture and the compact MobileNet network. Instead of utilizing a single CNN, some authors have proposed ensembles of CNNs for COVID-19 detection. Guarrasi et al.\cite{guar21} employed Pareto-based multiobjective optimization to build a CNN ensemble, and Rajaraman et al \cite{rajamaran21} showed that iterative pruning of the task-specific models not only improved prediction performance on the test data but also significantly reduced the number of trainable parameters.

Other authors have tried to optimize performance by designing a CNN architecture tailored to CXR classification. These models are inspired by existing architectures such as CoroNet \cite{iqbal}, which was inspired by the Xception design, DarkCovidNet \cite{ozturk}, which is based on the DarkNet \cite{redmon} CNN, and COVID-CAPS \cite{asfar}, which capitalizes on the capsule networks that preserve the spatial information between images. Instead of using an existing architecture, Wang et al. \cite{wang2020} employed generative synthesis to develop COVID-Net---a machine-designed deep CNN. An interesting approach combining CNN with graph CNN was presented by Kumar et al. \cite{kumar21} and achieved classification accuracy as high as 97\% on the COVIDx dataset. Other approaches rely on adding a class decomposition layer \cite{abbas21} to a pretrained CNN or a specific domain adaptation module to a fully convolutional network \cite{zhang20}.

In addition to using only CNN to classify chest CXR, specific features have been proposed and extracted to boost the classification performance \cite{varela}. Moreover, the extracted features can be combined with CNN to provide more robust prediction \cite{soda2021aiforcovid}.

Methodologies that do not focus on architecture improvements but rather attempt to improve pre- or postprocessing include the work of Heidari et al. on preprocessing X-ray images \cite{heidari} or Mor\'is et al. \cite{Moris21} on advanced data augmentation for improving COVID-19 screening.

Many of the above approaches have shown very promising results and achieved high classification accuracies. However, these methods must be considered with caution because several additional aspects must be considered before accepting a particular design as a production-ready solution. First, many of the aforementioned studies combined the COVID-19 dataset for experiments with other publicly available datasets to create a dataset used for model training and testing. This increases the chance that the model provides an output based on not only disease-related features but also dataset-specific aspects, such as contrast and saturation. Second, some studies, such as \cite{asfar} and \cite{wang2020}, utilized only simple hold-out model validation. Criticisms of previous studies are detailed in \cite{tabik}, where the authors propose a more robust solution called COVID-SDNet and utilize a new dataset for model validation. Biases in released datasets used to train diagnostic systems are also discussed in \cite{catala21}.

Some authors have considered other practical aspects of CXR classification, such as the limited available datasets. Oh et al. \cite{oh} proposed a solution for overcoming the lack of sufficient training datasets based on a pretrained ResNet-18, which processed CXR images through smaller patches. The authors of \cite{zhang2020viral} approached viral pneumonia detection as an anomaly detection problem. Using this approach, they were able to avoid training the model on large numbers of different pneumonia cases and focus solely on viral pneumonia. Recently, Luo et al. \cite{luo2020} proposed a framework to integrate the knowledge from different datasets and effectively trained a neural network to classify thoracic diseases.

To date, all previous approaches have relied on backbone networks pretrained on ImageNet. Transfer learning makes CNNs trained on large-scale natural images suitable for medical images. However, the disparities between natural images and X-ray images are quite significant. Training a CNN from scratch on a large X-ray dataset can further boost performance. Some early papers utilized self-supervised learning (SSL) \cite{sowrirajan2020},\cite{wangpr}, confirming that this is a viable approach. More recently, the authors of \cite{abbas21_trans} proposed a self-supervised approach guided by super sample decomposition and reported $99.8\%$ accuracy. Additionally, Rivero et al. \cite{rivero21} designed a graph-based deep semisupervised approach that needs only a very small labeled dataset and provides results competitive with supervised approaches.

COVID-19 detection from CXR images is a very hot research area, and new papers are appearing continuously. Covering all recent advances is outside the scope of this work. We have tried to mention different approaches and some representative cases of CXR classification; however, for more comprehensive reviews, we refer the interested reader to \cite{nabavi21}, \cite{ulhaq20}, \cite{bouch21}, \cite{shi21}, \cite{chowd20}, \cite{dong21}. These reviews summarize recent works and provide multiple perspectives on the most recent advances in COVID-19 detection from CXRs.

\subsection{Contrastive learning of visual representations}
Self-supervised neural networks provide unprecedented performance in computer vision tasks. Generative models operate mostly in the pixel space, which is computationally expensive and unsustainable on larger scales. On the other hand, contrastive discriminative methods operate on the augmented views of the same image, thus avoiding the computationally costly generation of the pixel space. In addition, contrastive discriminative methods currently achieve state-of-the-art performance on SSL tasks\cite{oord2018representation} \cite{henaff2020data}. Various approaches for self-supervised model training exist. The main paradigm has shifted towards instance discriminative models, where similar contrastive learning (SimCLR) \cite{chen2020simple}, momentum contrast for unsupervised visual representation learning (MoCo) \cite{he2020momentum} and bootstrap your own latent architecture (BYOL) \cite {grill2020bootstrap} have demonstrated as-yet-untapped potential. The representations learned by these architectures are on par with those of their supervised counterparts \cite{misra2020self} \cite{newell2020useful}.

From the point of view of pretext task selection, contrastive learning can be divided into context-instance contrast and context-context contrast \cite{liu2020self}. The former tries to find relations between the local features and the global representation of an instance (i.e., wheels and windows to a car). We believe that the learned local features help to distinguish between the target classes. Some examples of pretext tasks working in the context-instance principle are a jigsaw puzzle \cite{noroozi2016unsupervised} and rotation angle detection \cite{gidaris2018unsupervised}.

Context-context contrast architectures focus on the relationships between the global representations of different samples. CMC \cite{oord2018representation}, MoCo \cite{he2020momentum}, and SimCLR \cite{chen2020simple} contrast between positive and negative pairs, where the positive pairs constitute the same image augmented in different ways, while the negative pairs constitute all remaining images. The number of negative and positive pairs depends solely on the type of self-supervised architecture.

SimCLR and MoCo share the idea of using positive and negative pairs, but they differ in how the pairs are handled. In SimCLR, \cite{he2020momentum}, negative pairs are processed within the batch; thus, SimCLR requires a larger batch size. MoCo's representations of negative keys are maintained in a separate queue encoded by a momentum encoder.

BYOL claims to achieve better results than SimCLR and MoCo without using negative samples in its loss function. Different from SimCLR and MoCo, BYOL employs an $L_2$ error loss function instead of contrastive loss while using a principle similar to the momentum encoder introduced in MoCo.

BYOL takes advantage of two neural networks called ``online'' and ``target'' networks that learn by interactions between each other. BYOL initializes the optimization step by including one augmented view of a single image. It teaches the online network to correctly predict the representation of a differently augmented view of the same image produced by the target network.

\section{Methods}
\subsection{Self-supervised learning}
SSL is a subset of unsupervised learning methods that aim to learn meaningful representations from unlabeled data. The representations can then be reused for downstream (target) tasks as either a base for fine-tuning or as a fixed feature extractor for models such as logistic regression, SVM, and many others. Because manually annotated labels are not available in the training data, the SSL's first step is to generate pseudolabels automatically through carefully selected pretext tasks.

Formally, SSL can be defined as the minimization of an objective function $J(\boldsymbol{\theta})$ parameterized by parameters $\boldsymbol{\theta} \in \mathbb{R}^d$, which represents the mean loss over all training samples:
\begin{equation}
    J(\boldsymbol{\theta}) = \mathbb{E}_{\mbf{x} \sim  \hat{p}_{\rm {data}}} \mathcal{L} (m(\mbf{x}; \boldsymbol{\theta}), \pi(\mbf{x})),
\end{equation}
where $\hat{p}_{\rm{data}}$ is an empirical distribution, $\mathcal{L} $ is the per-example loss function, $m(\cdot , \cdot)$ is the model prediction when the input is $\mbf{x}$, and $\pi(.)$ is a function that returns pseudolabels for input $\mbf{x}$ based on the pretext task.

Optimization of such a neural network is accomplished similarly to supervised learning -- by updating the parameters $\boldsymbol{\theta}$ in the direction of the antigradient using methods based on stochastic gradient descent:

\begin{equation} 
\boldsymbol{\theta}^{(t+1)} = \boldsymbol{\theta}^{(t)} - \eta \frac{1}{B} \sum \limits_{i=Bt+1}^{B(t+1)} \frac{\partial{\mathcal{L} (m(\mbf{x}_i; \boldsymbol{\theta}), \pi(\mbf{x}_i))}}{\partial{\boldsymbol{\theta}}},
\end{equation}
where $\mathcal{L} $ is a loss function of the $i$-th example from the batch sampled at time $t$, $B$ stands for the batch size, and $\eta$ stands for a hyperparameter called the learning rate.

Modern SSL designs decouple the neural network architecture from downstream tasks, which makes the transfer of knowledge more straightforward. State-of-the art SSLs such as SimCLR \cite{chen2020simple} and MoCo \cite{he2020momentum} use the ResNet50 \cite{he2016deep} architecture on datasets such as CIFAR-10 \cite{cifar10} and ImageNet \cite{ILSVRC15} just as their supervised counterparts do.

\subsection*{Transfer learning}

Despite recent advances in deep learning and hardware accessibility, neural network training still tends to be slow and resource intensive. The transfer of knowledge from one domain to another reduces these burdens and has proved effective in numerous applications \cite{van2014transfer} \cite{shin2016deep}.

Transfer learning is applicable to tasks with different degrees of label availability. The knowledge extracted from a base domain can be acquired in an unsupervised, semisupervised, or supervised fashion. For unsupervised pretraining, transfer learning is defined as follows. Let $\mathcal{D}_S = (\mathcal{X}_S, \mathcal{P}_S)$ be a pretext dataset consisting of a set of samples $\mathcal{X}_S$ with corresponding pseudolabels $\mathcal{P}_S$ generated by the underlying pretext task and a downstream dataset $\mathcal{D}_T=(\mathcal{X}_T, \mathcal{Y}_T)$, where $\mathcal{X}_T$ denotes the set of training samples and $\mathcal{Y}_T$ denotes the set of true labels. Given example source datasets $\mathcal{D}_S$, pretext tasks $\mathcal{T}_S$, downstream datasets $\mathcal{D}_T$, and downstream tasks $\mathcal{T}_T$, transfer learning aims to reduce the loss function $\mathcal{L}$ of the model used for downstream tasks ($\mathcal{T}_T$) using the knowledge acquired from pretext tasks $\mathcal{T}_S$, where $\mathcal{D}_S \neq \mathcal{D}_T$ or $\mathcal{T}_S \neq \mathcal{T}_T$.

\subsection{Proposed approach}

We propose a neural network that solves pretext tasks based on contrastive instance discrimination similar to those used in SimCLR \cite{chen2020simple} and MoCo \cite{wu2018unsupervised}. The data augmentation module creates two versions of each image in the current batch, thus creating positive and negative pairs.

A positive pair is formed by two versions of one image that were augmented differently. Conversely, two images are denoted as a negative pair when they are augmented from different base images.

A neural network learns to solve this task by comparing representations of positive and negative pairs. It discriminates between them by maximizing the agreement between representations of two instances of the same image.

The representations are compared by cosine similarity, which is defined for two vectors $\mbf{u}, \mbf{v}$ as follows:
\begin{equation}
    \label{eq:similarity}
    \mathrm{sim}(\mbf{u}, \mbf{v}) = \frac{\mbf{u}^T \mbf{v}}{{||\mbf{u}||}_2 \cdot {||\mbf{v}||}_2}.
\end{equation}

Other similarity functions, such as Euclidean distance or dot product, could also be employed.

\begin{figure*}[h!]
    \centering
    \includegraphics[width=1\textwidth]{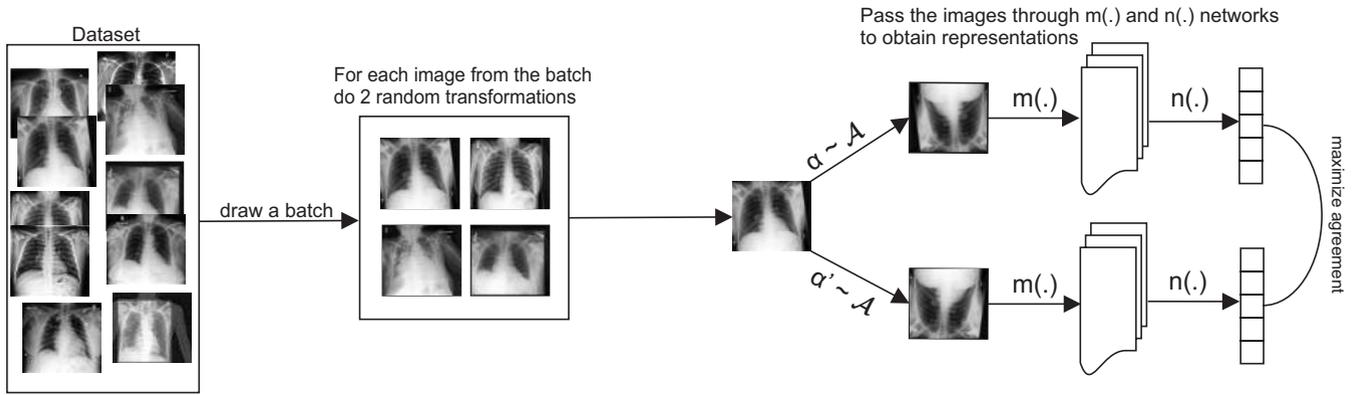}
\caption{Self-supervised architecture}
    \label{fig:simclr}
\end{figure*}

The proposed learning architecture consists of three parts: a backbone neural network $m(.)$, a projection head $n(.)$, and a stochastic data augmentation module $\mathcal{A}$. The backbone network $m(.)$ is a ResNet50 Wide network that extracts representations from the augmented data examples. The projection head $n(.)$ (see Table \ref{tab:projection_head}) transforms the output from the backbone network into a latent space where contrastive loss is applied. The size of the output vector is a hyperparameter allowing the final size to be adjusted to properly reflect the size of the original image. The data augmentation module $\mathcal{A}$ is a module that returns random augmentations as follows: resized crop, horizontal flip, rotation, Gaussian blur and color jitter. A resized crop involves a random crop of the image followed by a resize back up to the original image size. The entire learning architecture is depicted in Fig. \ref{fig:simclr}.

The optimization step is performed as follows. First, for a minibatch sampled at time t and consisting of images $\mathcal{X}_t = \{\mbf{x}_{Bt+1}, \mbf{x}_{Bt+2}, \ldots, \mbf{x}_{Bt+B}\}$ of size $B$ is drawn from the dataset samples $\mathcal{X}$, $\mathcal{X}_t \subseteq \mathcal{X}$, similar to supervised learning. Then, for each image in the minibatch, a positive pair is formed by augmenting the image twice with random augmentations, from which $2B$ images are obtained. The images are then encoded via the backbone network $m(.)$ to obtain the representation vectors. The representations are passed through the projection head $n(.)$ to obtain projection vectors. The set of projection vectors is denoted as $\mathcal{Z}_t$. To calculate the model error, we apply the NT-Xent loss (\textit{normalized temperature-scaled cross entropy loss}) introduced in \cite{sohn2016improved}. For a positive pair $(\mbf{z}_i,\mbf{z}_j)$ drawn from the set of projections of augmented images, $\mathcal{Z}_t$ is loss calculated as follows:

\begin{equation}
    \label{eq:infonce}
    l(\mbf{z}_i, \mbf{z}_j) = -\log \frac{e^{\mathrm{sim}(\mbf{z}_i, \mbf{z}_j)/\tau}}{\sum\limits_{z_k \in \mathcal{Z}_{t}- \{\mbf{z}_i\}} e^{\mathrm{sim}(\mbf{z}_i, \mbf{z}_k)/\tau}},
\end{equation}
where $\mathrm{sim}$ is a similarity function, $\mathcal{Z}_t - \{\mbf{z}_i\}$ are the $2B$ projections of augmented images (with the exception of the projection $\mbf{z}_i$), and $\tau$ is a hyperparameter called temperature.

After the loss is calculated, we backpropagate the errors to optimize the weights of the backbone neural network $m(.)$ and the projection head $n(.)$. At the end of the training process, we extract the features from the last layer of the backbone neural network $m(.)$ and discard the projection head $n(.)$.

The convergence criterion of the loss function can be approached similarly to supervised methods by looking at the validation loss curve. Most state-of-the-art SSL methods, such as MoCo \cite{wu2018unsupervised} and SimCLR \cite{chen2020simple}, do not employ an early stopper but set the number of epochs to a fixed number.

To mitigate the error from prolonged training and possible overfitting, a cosine annealing learning rate scheduler and weight decay are utilized.

\begin{table}
\caption{Projection head $n(.)$}
\centering
\begin{tabular}{|l|cc|}
\hline
Layer & Size & Bias \\
\hline
Global Average Pooling Layer& - & -  \\
Dense Layer & 2048 & True \\
Batch Normalization Layer \& ReLU & - & - \\
Dense Layer & 128 & False\\
\hline

\end{tabular}
\label{tab:projection_head}

\end{table}

\section{Data}

In this study, we utilized several CXR datasets. First, a large-scale dataset is required for network pretraining. Therefore, to formulate the pretext task, we utilized the CheXpert dataset \cite{irvin2019chexpert}, which contains 224,316 chest radiographs from 65,420 patients. The samples were labeled by extracting data from radiology reports from October 2002 to July 2017 for 14 commonly observed conditions, such as pneumonia, pneumothorax, and cardiomegaly. It should be noted that even though labels are available, we do not use these during pretraining because the proposed model is unsupervised.

To evaluate the transferability of the model to an external target dataset, we acquired four public datasets. First, the Cell dataset \cite{kermany2018identifying} comprises 5,323 
X-ray images from children, including 3,883 cases of viral and bacterial pneumonia and 1,349 normal images. The labels were provided by two expert physicians and verified by a third physician. The second dataset is the ChestX-ray14 \cite{wang2017chestx} dataset, which comprises 112,120 X-ray images with eight disease labels from 30,805 patients. We used only a subset of this dataset by selecting only patients with pneumonia and a matched number of healthy controls. The other two datasets were compiled only recently and were intended for COVID-19 detection. The C19-Cohen dataset \cite{cohen2020covidProspective} (accessed 21.10.2020) is a collection of different types of pneumonia (viral, bacterial, and fungal). We selected two classes: 304 patients with COVID-19 and 114 patients with other types of pneumonia. Finally, we also evaluated the proposed model on the COVIDGR dataset \cite{tabik}, which contains 426 CXR images of patients with COVID-19 of four different severity levels and the same number of control subjects. Note that while 76 of these 426 COVID-19 patients were diagnosed as positive by PCR, their CXRs were evaluated as normal, making the classification task more challenging. A brief summary of all the datasets utilized in this study is presented in Tab. \ref{tab:data}.

\begin{table}
\caption{Datasets used in this study}
\centering
\begin{tabular}{|lcccc|}
\hline
Dataset & \# samples &\# class 0 & \# class 1 & source \\
\hline
\multicolumn{5}{c}{pretext}\\
\hline
CheXpert& 224,316 & na & na & \cite{irvin2019chexpert} \\
\hline
\multicolumn{5}{c}{target}\\
\hline
Cell & 5,323 & 3,883 & 1,349 & \cite{kermany2018identifying} \\
ChestX-ray-14 &  2,706 & 1,353 & 1,353 & \cite{wang2017chestx}\\
C19-Cohen & 807 & 564 & 243 & \cite{cohen2020covidProspective}\\
COVIDGR & 852 & 426 & 426 & \cite{tabik}\\
\hline

\end{tabular}
\label{tab:data}

\end{table}

\section{Experiments and results}

In this section, we analyze some aspects of the network pretraining task and report the experimental results.

To demonstrate the generalizability of our approach, we formulated four classification tasks on four publicly available CXR datasets. We avoided combining datasets for a particular classification task and used only one dataset for a specific classification task. In this manner, we tried to avoid the bias and criticism outlined in \cite{tabik}. The datasets details are presented in Tab. \ref{tab:data}. For the Cell dataset and ChestX-ray-14, we classified subjects as having pneumonia or as healthy. Similarly, for COVIDGR, we discriminated between patients with and without COVID-19 disease. Finally, because the C19-Cohen dataset does not include healthy controls, we discriminated between COVID-19 and other types of pneumonia.

\subsection{Network training and feature extraction}
\label{sec:train}

\begin{figure}
    \centering
\hspace*{-2.0em}
   \includegraphics[width=0.55\textwidth]{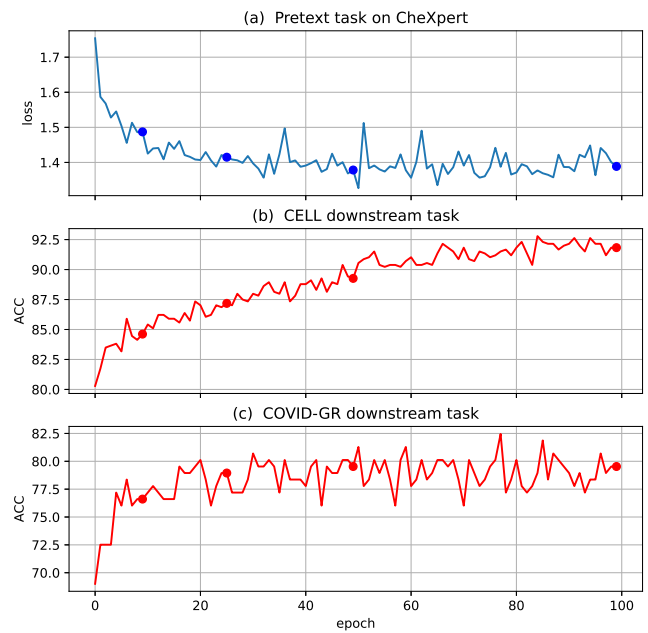}
\caption{ Loss and accuracy in each epoch. (a) Loss on the pretext task and accuracy on the 
(b) Cell dataset and (b) 
COVIDGR dataset.}
    \label{fig:training}
\end{figure}

We pretrained a ResNet-50 Wide model in the self-supervised task-agnostic way on the large CheXpert dataset of CXR images. The effective batch size was set to 128, and the temperature value was 0.5. We used the Adam optimizer \cite{kingma2014adam} with a learning rate of 0.0005. The model was trained for 100 epochs without any stopping criteria. We used cosine annealing  to change the learning rate during the training. The training process convergence is depicted in Fig. \ref{fig:training} (a). The loss on the pretext task plateaus at approximately the 25th epoch and does not decrease further but instead oscillates around some value. This result raises the question of whether it is necessary to train the model beyond the 25th epoch. However, the relationship between the loss on the pretext task and the model's performance on the target task has not yet been established. Here, we analyze the performance in terms of the prediction accuracy on the two datasets (the Cell dataset and the COVIDGR dataset). The prediction accuracy for each epoch is depicted in Fig. \ref{fig:training} (b) and Fig. \ref{fig:training} (c). Although the loss on the pretext task does not improve significantly beyond the 25th--30th epochs, the accuracy achieved for predictions on the Cell dataset increases when we employ models that have been trained with a larger number of epochs. This indicates an interesting phenomenon. During pretraining, even after the loss on the pretext task is no longer improving, the model is still learning. In contrast to the prediction accuracy obtained on the Cell dataset, for the COVIDGR dataset, the accuracy does not gradually improve for models trained beyond the 25th epoch. However, this can be explained by the composition of the images in the dataset. As explained before, COVIDGR also contains CXR images from COVID-19 patients in which expert radiologists did not find any pathological changes. The highest reported result so far on this dataset is $76.18 \%$ \cite{tabik}. Our model achieved this accuracy level quite early; thus, we hypothesize that there was no room for further improvement.

\begin{figure*}
    \centering
    \includegraphics[width=1\textwidth]{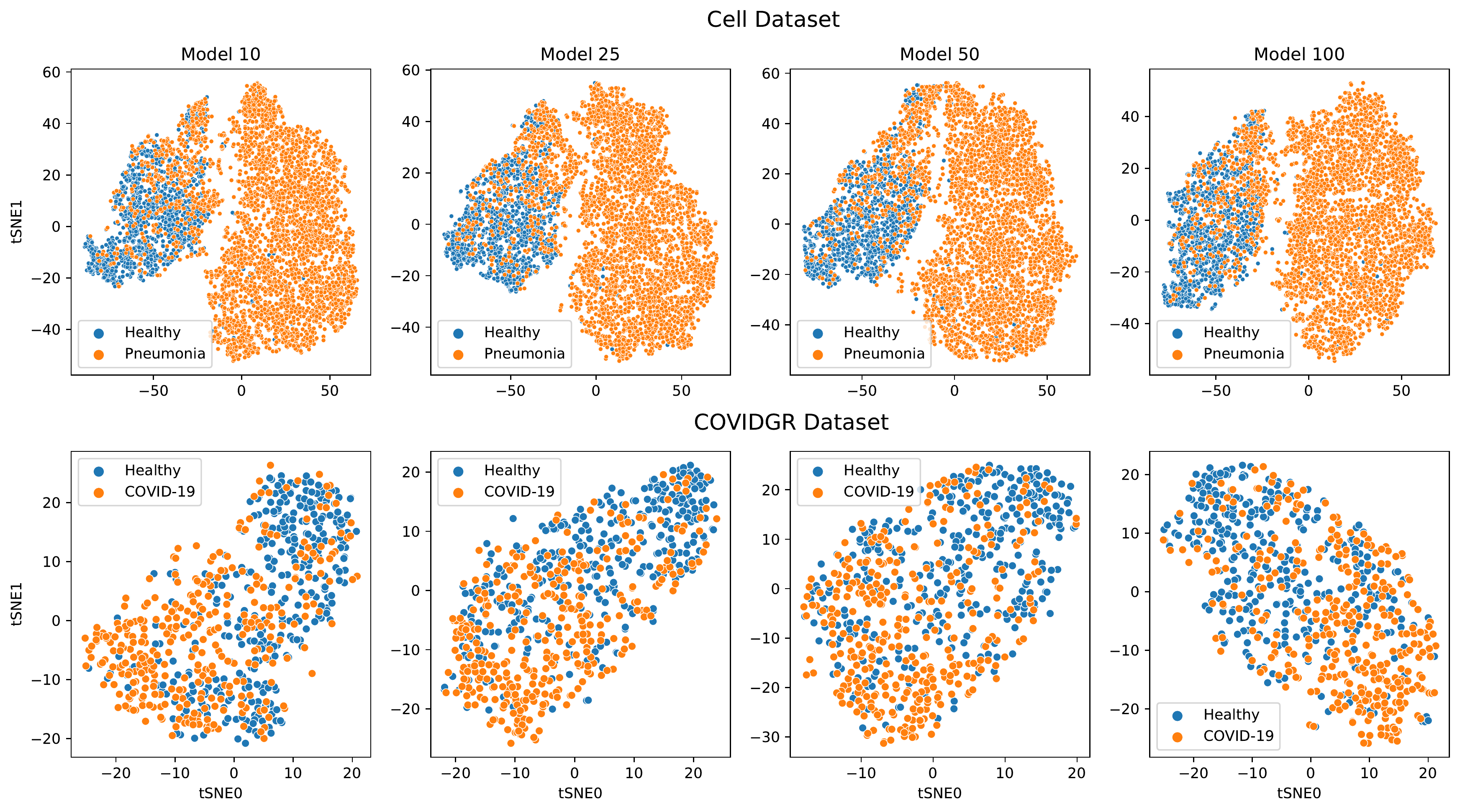}
\caption{t-SNE visualization of the features extracted by models selected in different stages of the pretraining process (10th, 25th, 50th and 100th epochs). The features shown are from the Cell and COVIDGR datasets.}
    \label{fig:tsne}
\end{figure*}

To visualize the learned representations, we chose models from four different checkpoints and extracted features. The models were trained for 10, 25, 50 and 100 epochs. Fig. \ref{fig:tsne} shows t-SNE visualizations of features extracted from the Cell and COVIDGR datasets by these four models. While the Cell dataset exhibits a noticeable but slight improvement in the separability of two classes, the two classes for the COVIDGR dataset seem to be interlaced through all four images.

\subsection{Numerical results}

To examine the predictive performance of the proposed approach, we employ transfer learning and use the pretrained network as a fixed feature extractor. The size of the extracted feature vector is determined by the last dense layer in the projection head, which was 128 in our network. We adopted logistic regression as a classifier and evaluated the model on four different CXR datasets. We also evaluated other linear classifiers, but the results and general trend were very similar, so we omit those results for the sake of readability. To ensure the model's generalizability and avoid overfitting, we used stratified 5-fold cross-validation. The datasets were divided into training, validation, and testing subsets. If the original paper that introduced the datasets also provided the specification of training/validation/test subsets of data, we used that division to achieve fair comparisons of model accuracy with the published results (COVIDGR, ChestX-ray-14 and Cell). Otherwise, we divided the data as follows: 70\% as training samples, 10\% for validation and 20\% as test data (C19-Cohen). Furthermore, we ensured that the CXR images of a particular patient were present only within the same data subset to prevent data leakage that would cause positive bias. All CXR images were resized to 224x224 pixels.

To determine the optimal logistic regression parameters, we searched through the parameter space $\mathrm{C}=\{0.01, 0.05, 0.1, 0.2, 0.5, 1\}$. The logistic regression weights were automatically adjusted to be inversely proportional to the class frequencies in the input data. The other parameters were set to their default values. The best model was adopted after a grid search based on the area under the receiver operating characteristic curve (AUC) metric.

We also evaluated the amount of data required for the pretext task to correctly identify relevant features that are beneficial for the downstream task by testing with three different data fractions: $1\%$, $10\%$ and $100\%$.

The results of the trained models are depicted in Table \ref{tab:results_chestxray}. To provide a better overview of model performance, we calculated several metrics: accuracy (ACC), AUC, sensitivity (SEN) and specificity (SPE).

Our first observation is that the prediction accuracy differs between the datasets. The model prediction accuracy varies more in the pneumonia classification task (Cell, ChestX-ray-14) than in the COVID-19 classification task (C19-Cohen, COVIDGR). This variation is caused by the different dataset compositions. The datasets were compiled from different sources and acquired by different devices, which influences the image characteristics. However, more importantly, the Cell dataset is composed of CXR children aged four to six years, making classification a specific type of task.

On the Cell dataset, the AUC and ACC tend to increase as the dataset fraction size increases for the pretext task. The highest AUC $97.7\%$ of our model is higher than the $96.6 \%$ reported in \cite{kermany2018identifying}, which used transfer learning based on ImageNet. This result shows that representations learned in a self-supervised fashion on smaller datasets with semantically closer domains are more beneficial than supervised pretraining on large but semantically very different datasets such as ImageNet.

The model evaluated on the ChestX-ray-14 dataset yielded significantly lower results than the model evaluated on the Cell dataset. In this case, 
the results are independent 
of the fraction of the dataset used for training the pretext task. Tab. \ref{tab:results_chestxray} shows that the proposed model achieves a higher score than the results published in \cite{wang2017chestx}. However, the comparison is not entirely fair because the authors of \cite{wang2017chestx} were solving a multiclass problem, which could have had a negative impact on the model accuracy compared to binary classification.

Models trained on the COVIDGR and Cohen-19 datasets achieved comparable results. One model trained on C19-COHEN achieved AUCs up to 91.5\% when using a 10\% fraction of the dataset in the pretext task. Surprisingly, it outperformed another model trained on the pretext task with the entire dataset. We hypothesize that this may have been caused by the better performance of the logistic regression model due to the hyperparameters found in the grid search. Some of the previously published papers combined the C19-Cohen dataset with CXRs of healthy controls obtained from other datasets. We intentionally avoided such combinations, and to the best of our knowledge, no published paper has conducted classification only on the C19-Cohen dataset; thus, we cannot directly compare the performance of our model with those of others on the C19-Cohen dataset.

Encouraging results were achieved on the COVIDGR dataset in differentiating between healthy and COVID-19 CXRs. Our CNN pretrained in a self-supervised fashion was able to outperform the supervised COVID-SDNet \cite{tabik} model by a few percent. Although this difference is not large, it should be noted that SSL does not require a large, labeled training dataset, which can save substantial human resources.

\begin{table*}[h]
    \centering
    \caption{Prediction performance of the proposed approach on four CXR datasets. We provide previous published results for comparison purposes. * There is no available result for comparison because previously published results combine the C19-Cohen dataset with some other CXR dataset. Moreover, data are continuously added to the 19-Cohen dataset. ** Result of pneumonia prediction as part of a multiclass classification. }
    \hspace*{-2.8mm}
    \scalebox{0.92}{
    \begin{tabular}{@{}|lccccccccccccccccc|@{}}
    \hline
    \multicolumn{1}{|c|}{\multirow{3}{*}{Dataset}} & 
    \multicolumn{12}{c|}{Fraction of dataset} & \multicolumn{4}{c|}{ImageNet pretraining} & Published result \\ 
    \hline 
    \multicolumn{1}{|c|}{} & \multicolumn{4}{c|}{1\%} & \multicolumn{4}{c|}{10\%} & \multicolumn{4}{c|}{100\%} & \multicolumn{4}{c|}{} & \multirow{2}{*}{ACC/AUC} \\ 
    \multicolumn{1}{|c|}{}  & \multicolumn{1}{l|}{ACC} & \multicolumn{1}{l|}{AUC} & \multicolumn{1}{l|}{SEN} & \multicolumn{1}{l|}{SPE}  & \multicolumn{1}{l|}{ACC} & \multicolumn{1}{l|}{AUC} & \multicolumn{1}{l|}{SEN} & \multicolumn{1}{l|}{SPE} & \multicolumn{1}{l|}{ACC} & \multicolumn{1}{l|}{AUC} & \multicolumn{1}{l|}{SEN} & \multicolumn{1}{l|}{SPE}  & \multicolumn{1}{l|}{ACC} & \multicolumn{1}{l|}{AUC} & \multicolumn{1}{l|}{SEN} & \multicolumn{1}{l|}{SPE} &\\ \hline
    Cell & 85.6 & 96.6  & 99.5 & 62.4 &  86.9 & 96.9 & 99.2 & 66.2  & 91.5& 97.7 & 98.7 & 79.5 & 83.2 & 94.9& 98.2 & 58.1 &  92.8/96.8 \cite{kermany2018identifying} \\
    C19-Cohen  & 84.9 & 89.2 & 90.6  & 73.6  & 81.8 & 91.5 & 85.8 & 73.6 &  81.1 & 88.2 & 83  & 77.4  &  69.8 & 77.0 & 80.2 & 49.1 &  na*\\
    COVIDGR & 79.5 & 86.6   & 83.5 & 75.6 &  77.8 & 86.0 & 80 & 75.6 &  78.4 & 87.1 & 83.5 & 73.3 &  71.3 &  77.7 & 69.4 & 73.2 & 76.16/na \cite{tabik}\\
    ChestX-ray-14& 71.5 & 79.1 & 72.8 & 71.4 & 71.2 & 78.1 & 72.0 & 71.2 &  71.4 & 78.4 & 71.3 & 71.5& 69.6 & 75.1 & 67.0 & 69.6 &  na/65.8** \cite{wang2017chestx}\\ \hline
    \end{tabular}
    }
    
    \label{tab:results_chestxray}
\end{table*}


\subsection{Explaining CNN decisions}

\begin{figure*}[h!]
    \centering
    \includegraphics[width=1\textwidth]{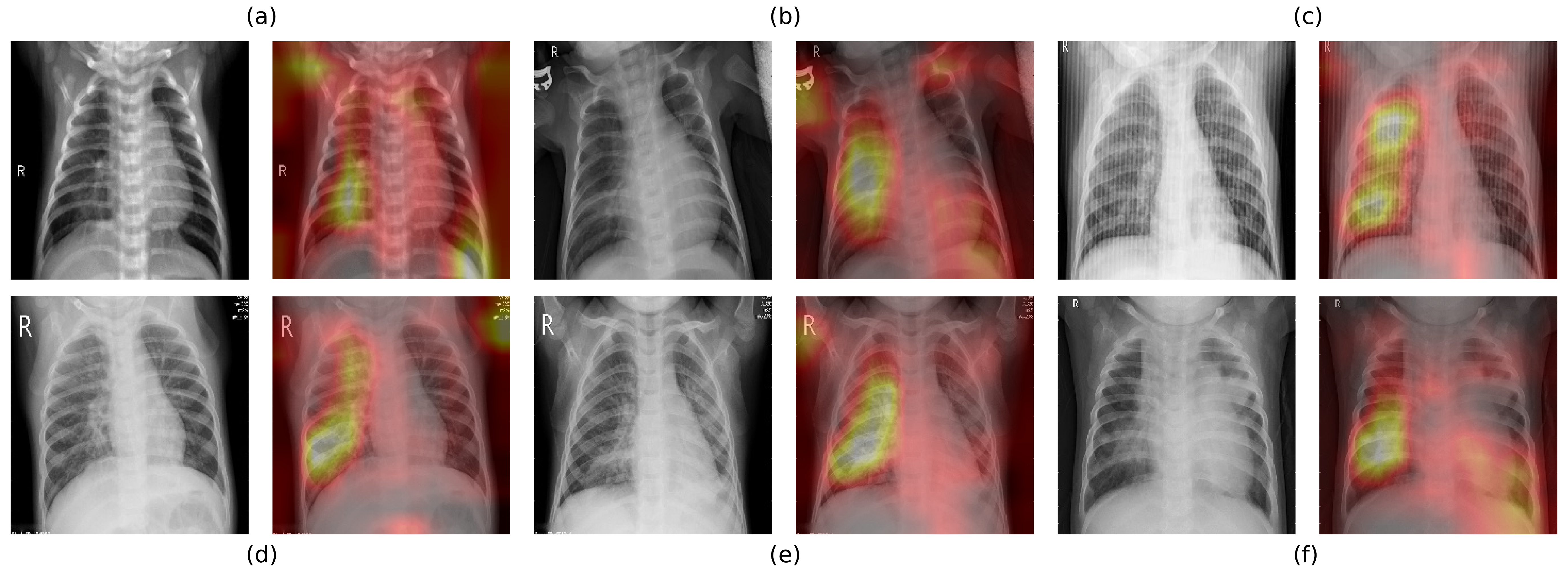}
\caption{Visualizations of 6 CXR images from the Cell dataset and their Grad-CAM attention maps. These images were taken of patients with pneumonia and were diagnosed with pneumonia by our model.}
    \label{fig:vis}
\end{figure*}

To shed some light on the CNN decisions, we employ gradient-weighted class activation mapping (Grad-CAM) to highlight the important regions of the CXR image corresponding to a decision. Grad-CAM is a class-discriminative localization technique that generates visual explanations for CNN decisions \cite{selvaraju2017grad}. Fig. \ref{fig:vis} shows CXRs of six different patients correctly classified as pneumonia cases. Here, images of both ground-glass opacities and consolidations are present together with air bronchograms. An air bronchogram is a dark radiographic appearance of an air-filled bronchus (dark thread-like line) made visible by the opacification of the surrounding alveoli (``white lung field''). An air bronchogram is another pathological sign frequently associated with pneumonia. The more edematous the lung tissue is, the easier it is to spot; however, this applies only to those bronchi  that have at least some residual air inside them. In this instance, the areas in the lungs highlighted by the attention map cover the regions with visible pulmonary opacification and air bronchograms, which provides the grounds for a correct diagnosis. In clinical practice, a radiologist looks for the same radiological signs as the neural network used here for decision making. Other areas highlighted by the attention map are outside the lung region. These areas cannot be linked to any pathology caused by pneumonia and probably reflect zones incorrectly evaluated by the visualization algorithm or by the CNN itself.

\section{Discussion}

We proposed an approach based on a self-supervised convolutional network and evaluated it on four datasets. To avoid the critiques presented by \cite{tabik} and others \cite{maguolo2020critic}, we did not combine existing datasets, which clearly limits the available training and testing data size. On the other hand, it helps ensure that the model learns only parameters related to specific aspects of the disease pathology and not differences arising from different devices or acquisition procedures. Here, it should be noted that we avoided combining datasets from different sources in that the disease class samples were provided by one source and the control subject samples were extracted from different sources. To increase the confidence of our approach, we evaluated the trained model on four different datasets focused on two different diseases. We differentiated between CXR with no findings and CXR-containing pathologies and between CXRs of patients with different pneumonia types and those with COVID-19-induced pneumonia.

One issue we touched on briefly in Section \ref{sec:train} is the selection of the optimal model for transfer learning. Fig. \ref{fig:training} clearly shows that the relationship between model performance on the pretext task and the subsequent downstream task is not straightforward. The question that arises is how to select the optimal checkpoint for the pretrained model. According to the traditional training view, the user may be tempted to stop the training at approximately the 40th epoch based on the loss curve in Fig. \ref{fig:training}(a) because the loss is no longer improving. However, at this point, the model is far from optimal in the sense of the prediction accuracy on the downstream task (at least for the Cell dataset \ref{fig:training}(b)). Further research is needed to establish the relationship between model performance on pretext tasks and downstream tasks.

\subsection{Limitations of the study}

We provide decision support for the classification of CXR images; however, the output should be taken with caution. The final diagnosis should always be based on the combination of clinical symptoms (cough, fever, pathological respiratory sounds) and laboratory results. However, in critical pandemic situations such as the one we are experiencing currently, medical staff are extremely busy, and a solution that provides rapid automated information could help to reduce the burden on medical personnel.

Some datasets, such as C19-Cohen, are compiled from X-ray images from different institutions, so this could result in some bias. However, the images from different resources are distributed in both classes, so potential bias should be quite limited. This is a dataset frequently utilized in similar studies, so we included it for comparison even though the results should be interpreted carefully.

Clearly, there is a strong need for further validation and detailed assessment led by transparent reporting of a multivariable prediction model for individual prognosis or diagnosis (TRIPOD) \cite{collins2019reporting} before an automated approach can be used in clinical practice. Additional datasets from different types of devices need to be included in testing and evaluation. However, this study has proven that it is possible to train the model in a self-supervised fashion and apply it successfully to medical imaging tasks without the need for large amounts of labeled data. This may open new research horizons because similar approaches can be examined for other types of medical imaging, such as computer tomography and retinal imaging.

The proposed approach for CXR classification is based on a deep CNN, and a fundamental principle of these models is that they work in a black-box manner. The ability to explain the decisions of deep learning models is still in its early stages and is a hot research topic. In this paper, we utilize grad-CAM to provide not only prediction but also some reasoning for the network decision. Grad-CAM shows the region that is most relevant for the prediction. The explanations are based on the learned attention regions in Fig. \ref{fig:vis} and are the same as areas that doctors review. However, these explanations do not cover all aspects and peculiarities included in the final decision. Because explainability is of crucial importance for medical applications, this and other limitations will be addressed in future work.

\subsection{Fault analysis}

\begin{figure}[h!]
    \centering
    \includegraphics[width=0.5\textwidth]{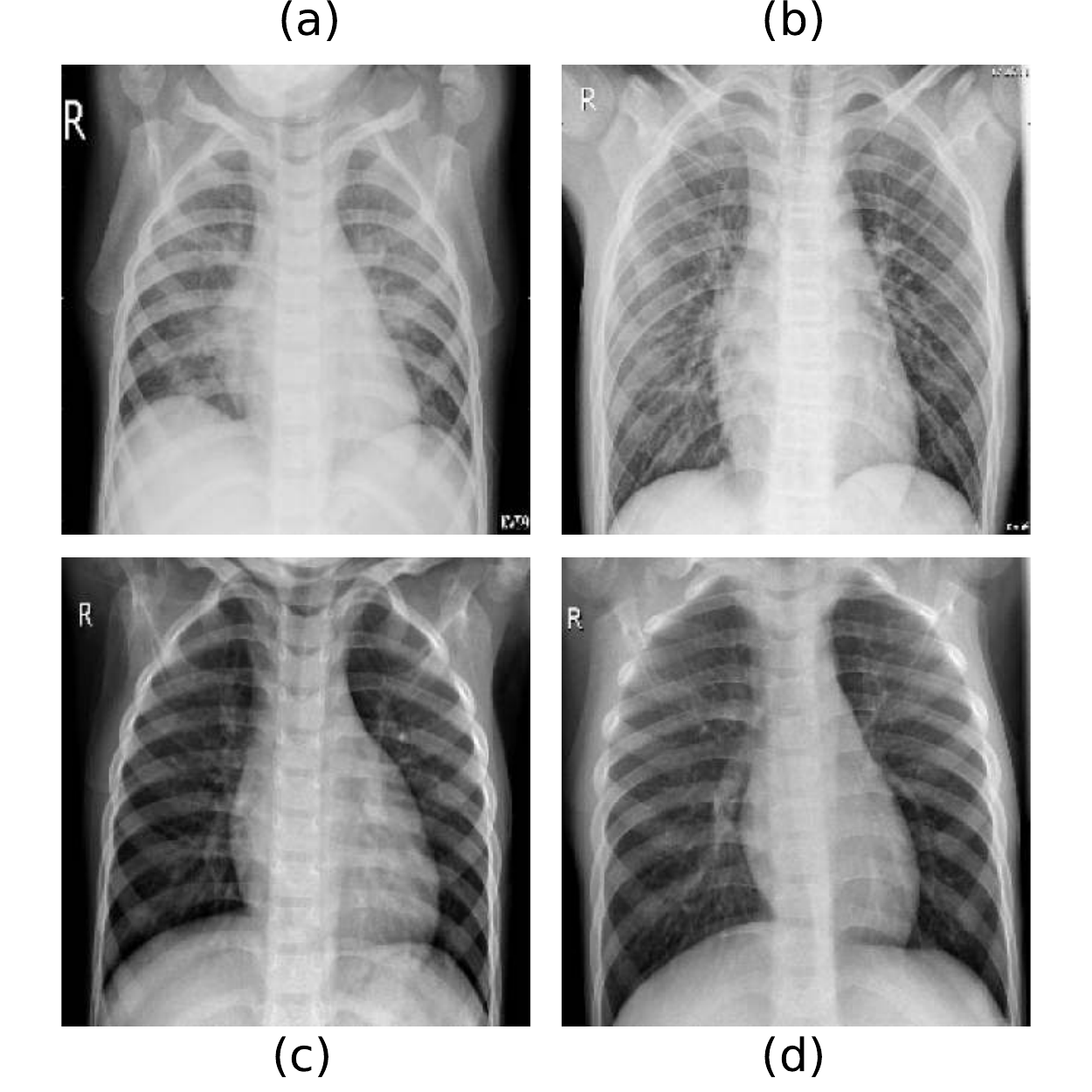}
\caption{CXR images of four healthy subjects classified as pneumonia.}
    \label{fig:faults}
\end{figure}

We also investigated some misclassifications to obtain a better understanding of CNN decisions. Fig. \ref{fig:faults} shows four CXRs that were incorrectly classified. Cases (a) and (b) were both misclassified as pneumonia. The patient in Fig. \ref{fig:faults}(a) has a small consolidation-like area (``white lung field'') in the (right) middle lobe, and Case (b) shows a distinct diffuse reticular interstitial pattern. Both patterns may resemble pneumonia findings, which may have led the model to incorrect classification. It is likely that a radiology specialist would make the same mistake if the only available information was the radiograph. The true cause of these patterns would have to be determined by the patient's clinical history and additional radiological examinations (such as computed tomography scan). On the other hand, the CXR images in \ref{fig:faults}(c) and \ref{fig:faults}(d) were also misclassified as pneumonia, but these do not present any structural changes that could be associated with the incorrect classification and should have been classified as healthy patients. It is difficult to determine the cause of these particular misinterpretations. This demonstrates the disadvantage of a CNN behaving as a black box.

\section{Conclusions}

The current pandemic further highlights the need to include diagnostic decision support systems in clinical decision making. The successful incorporation of these systems into contemporary medical devices could automate certain tasks and reduce medical personnel workloads. Automated solutions also contribute strongly during noncritical times by giving medical specialists more time for tasks and duties that require a more careful or specific approach.

As a contribution to medical expert systems, we introduced a solution that classifies CXR images. The proposed approach utilizes a CNN pretrained on an unlabeled dataset of CXR images. By avoiding the need for labeled data, which are both scarce and expensive in the medical domain, our approach offers new possibilities for CNN utilization by demonstrating that CNN networks do not need to be trained on only natural images (such as the ImageNet dataset), as in the majority of approaches today; they can instead be trained on images that are semantically closer to the target task. In our case, a network pretrained on the ChestXpert dataset was able to learn meaningful representations and extract relevant features for pneumonia and COVID-19 detection. The obtained results of our unsupervised model are competitive with their supervised counterparts. Considering that self-supervised contrastive learning for visual representations is a very new topic, this approach has huge potential. Later methodological improvements may further boost the performance.

\section*{Acknowledgment}

We would like to thank Dr. J. Bu\v{s}a and D. Hub\'a\v{c}ek, MD, for their valuable comments. \\

\bibliographystyle{IEEEtran}
\bibliography{unsupXRAY.txt}

\begin{IEEEbiography}
[{\includegraphics[width=1in,height=1.25in,clip,keepaspectratio]{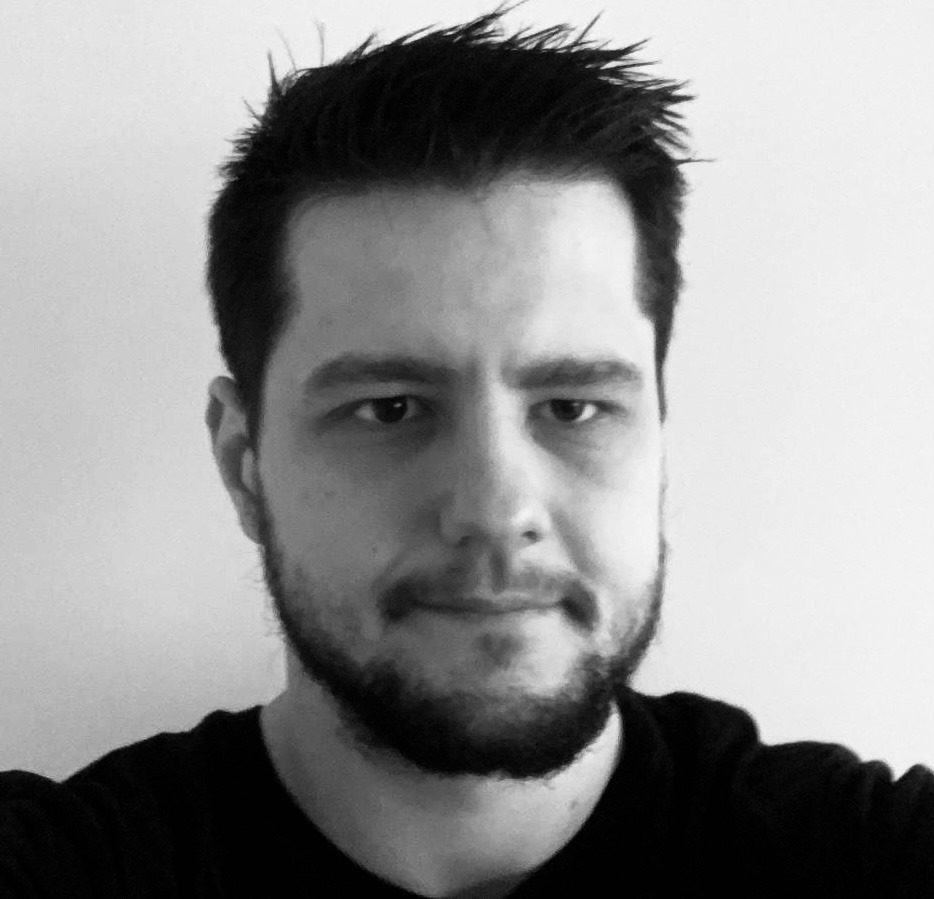}}]
{Matej Gazda} received an MSc degree in applied
computer science from the Faculty of Electrical
Engineering and Informatics, Technical University
of Kosice, Kosice, Slovakia, in 2019.
His research interests include biomedical image
analysis, feature selection, and biomedical decision
support systems.
\end{IEEEbiography}

\begin{IEEEbiography}
[{\includegraphics[width=1in,height=1.25in,clip,keepaspectratio]{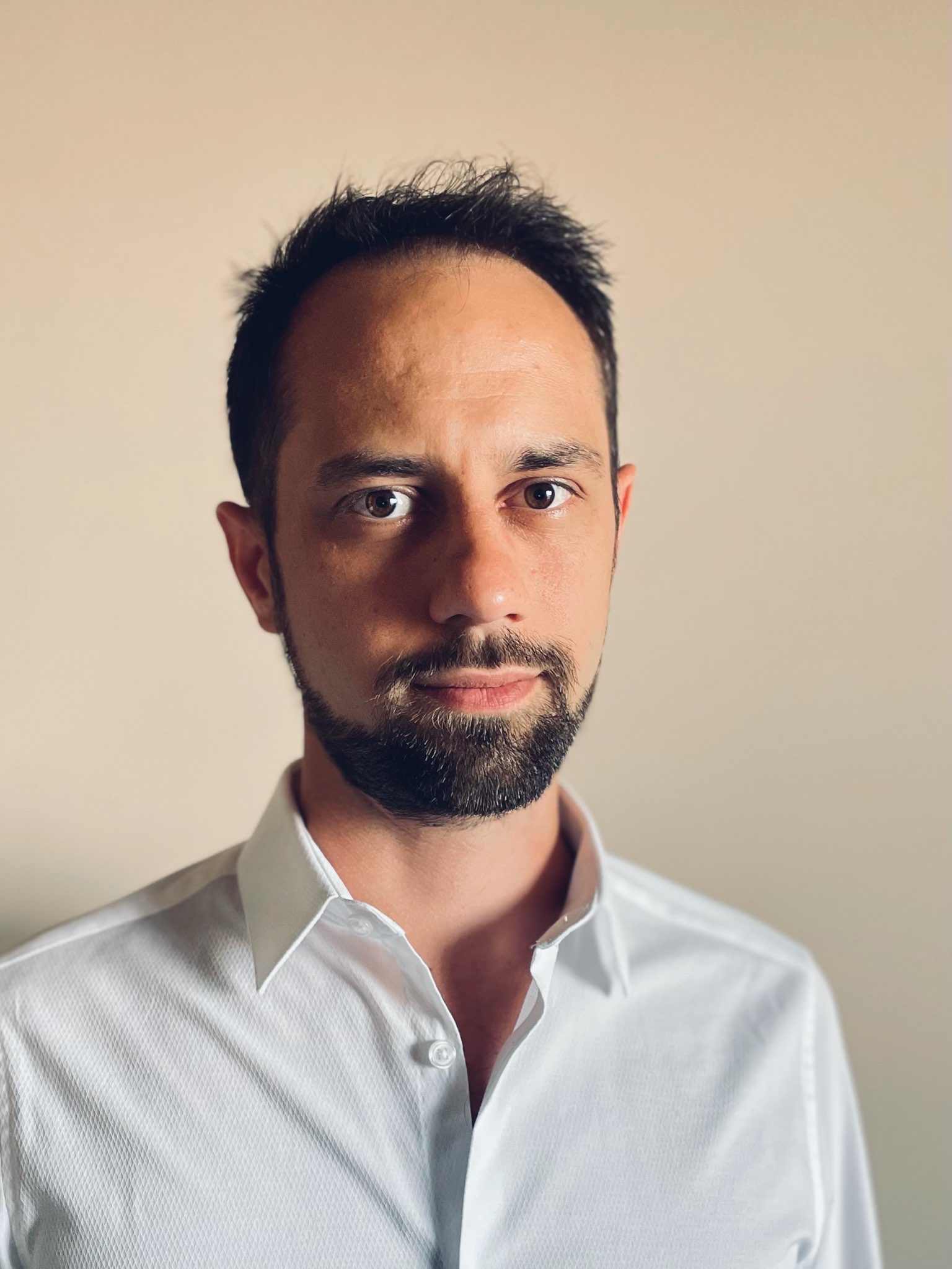}}]
{Jakub Gazda} received an MD degree in 2018 from the Medical Faculty of P. J. \v{S}afarik University, Kosice, Slovakia. He works at the Internal Medicine Department, and his research interests include autoimmune liver diseases and medical imaging.
\end{IEEEbiography}

\begin{IEEEbiography}
[{\includegraphics[width=1in,height=1.25in,clip,keepaspectratio]{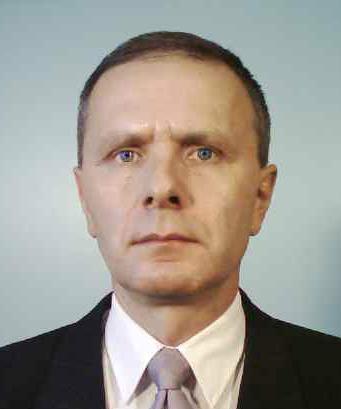}}]
{J\'an Plavka} graduated in discrete mathematics from P. J. \v{S}afarik University in Kosice. He defended his PhD in the field of steady-state discrete event dynamic systems in 1991. At present, he is a professor in the Department of Mathematics and Theoretical Informatics, Faculty of Electrical Engineering and Informatics, Technical University of Kosice. His scientific research focuses on computer science, the complexity of algorithms and discrete event dynamic systems. In addition, he investigates questions related to neuron networks.

\end{IEEEbiography}

\begin{IEEEbiography}
[{\includegraphics[width=1in,height=1.25in,clip,keepaspectratio]{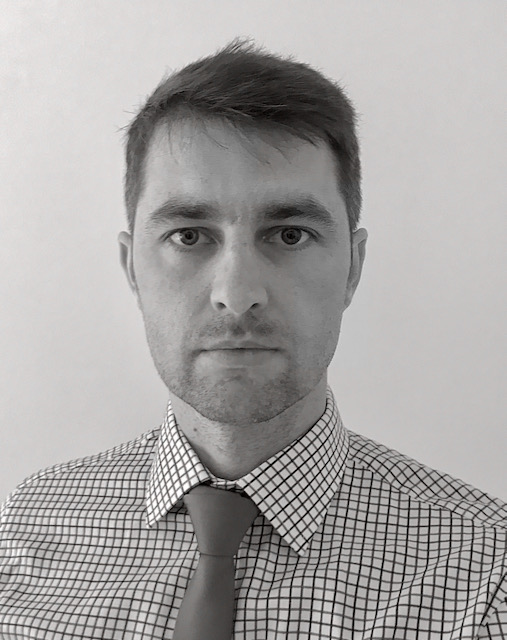}}]
{Peter~Drotar} (Member,  IEEE) received MSc and PhD degrees in electronics from the Faculty of Electrical Engineering and Informatics, Technical University of Kosice, Kosice, Slovakia, in 2007 and 2010, respectively. From 2010 to 2012, he was with Honeywell International, Advanced Technology Europe, as a Scientist for Communication and Surveillance Systems. From 2012 to 2015, he was with SIX Research Centre, Brno University of Technology, Brno, Czech Republic, as a Postdoctoral Research Assistant. He is currently an Associate Professor with the Department of Computers and Informatics, Technical University of Kosice. He leads research and development projects concerning biomedical decision support systems. His research interests include biomedical signal and image processing, feature selection and pattern recognition.
\end{IEEEbiography}
\EOD
\end{document}